\documentclass[thmsa,a4paper,onecolumn,11pt]{article}
\usepackage{amssymb}

\usepackage{sw20lart}

\input{tcilatex}
\begin{document}

\title{Information and The Brukner-Zeilinger Interpretation of Quantum Mechanics: A
Critical Investigation}
\author{A. Shafiee\thanks{%
E-mail: shafiee@sharif.edu} $^{\text{(1)}}$, \ F. Safinejad$^{\text{(2)}%
}\bigskip $ and F. Naqsh$^{\text{(3)}}$ \\
%EndAName
{\small \ }$\stackrel{1)}{}$ {\small Department of Chemistry, Sharif
University of Technology,}\\
{\small \ P.O.Box 11365-9516, Tehran, Iran}\\
{\small \ }$\stackrel{2)}{}$ {\small Department of Chemistry, Tarbiat
Modarres University,}\\
{\small \ P.O.Box 14115-175, Tehran, Iran.}\\
{\small \ }$\stackrel{3)}{}$ {\small Department of Chemistry, Kashan
University,}\\
{\small \ Kashan, 87317-51167, Iran.}}
\maketitle

\begin{abstract}
In Brukner and Zeilinger's interpretation of quantum mechanics, information
is introduced as the most fundamental notion and the finiteness of
information is considered as an essential feature of quantum systems. They
also define a new measure of information which is inherently different from
the Shannon information and try to show that the latter is not useful in
defining the information content in a quantum object.

Here, we show that there are serious problems in their approach which make
their efforts unsatisfactory. The finiteness of information does not explain
how objective results appear in experiments and what an instantaneous change
in the so-called information vector (or catalog of knowledge) really means
during the measurement. On the other hand, Brukner and Zeilinger's
definition of a new measure of information may lose its significance, when
the spin measurement of an elementary system is treated realistically.
Hence, the sum of the individual measures of information may not be a
conserved value in real experiments.
\end{abstract}

\section{Introduction}

Brukner and Zeilinger in a series of papers represent a new
information-based interpretation of quantum mechanics which its foundation
is established on the informative propositions [1-8]. They emphasize that
information is the most fundamental notion in quantum mechanics and the
other elementary concepts like randomness and entanglement are secondary in
quantum mechanics, because all of these concepts can be deduced from
definite rules of information.

Brukner and Zeilinger also believe that Shannon information\footnote{%
We use the statements like Shannon information, Shannon's measure of
information, Shannon's measure of uncertainty or Shannon entropy
interchangeably here.} [9] cannot be considered as an adequate measure of
uncertainty for quantum measurements, since it can be derived considering
the classical requirements. Unlike Shannon's approach, they do not think
about any communication channel for transmission of information; rather
their discussion concentrates on the information content of an isolated
system. Correspondingly, they define a measure of information which its
value depends on the experimental context. Yet, the \textit{total}
information is a major concept in their interpretation which is defined as
the total knowledge that an \textit{experimentalist} possesses before a
complete set of mutually complementary experiments are performed [2, 4]. In
contrast to the Shannon information, the \textit{total} information of a
quantum system is invariant under the change of a transformation from one
complete set of complementary variables to another. This means that the
total information does not change when one observes the quantum phenomena in
different ways [4].

The new measure of information introduced by Brukner and Zeilinger, can be
applied to describe the information content of a quantum system in various
situations. The entanglement swapping procedure indicated in [7] is an
example. In some way, it is used to express the necessary and sufficient
conditions for violation of a Bell inequality in an information-theoretical
language. Other authors used the measure in the context of state estimation
[10] or quantum random access codes [11].

Nevertheless, some critiques have also been appeared in different papers
[12-16]. The main purpose in all these works is to show that Shannon entropy
is an adequate and consistent measure of information in both the classical
and quantum regimes (in the latter case, in the form of von Neumann
entropy). So, some persons believe that there is no need to define an
alternative measure of information. But, these engaging criticisms do not
explain why the Brukner-Zeilinger measure of information cannot be regarded
as a proper definition in quantum domain. Furthermore, to the present
authors' knowledge, it has been remarked up to now no concrete criticism on
the very conceptual basis of their interpretation.

In Brukner and Zeilinger's interpretation, two mutually connected themes can
be identified. The first is based on this attitude that quantum systems
carry finite information and that the finiteness of information is a
fundamental principle in \textit{Nature} which is inherently irreducible.
So, it is expected that anything in the quantum domain can be either
extracted from or explained by the notion of information. The other theme is
that any quantum event can be described by a specific definition of a
measure of information. This allows one to represent the information content
of a quantum system via a vector in the space of information which changes
with a corresponding change of the experimental setup, but its length is
conserved. This means in turn that the total information content of a
quantum system is conserved and does not change under a transformation in
the information space.

In this paper we are going to analyze the key issues of Brukner and
Zeilinger's interpretation of quantum mechanics. In section 2, the basic
elements of their interpretation are reviewed. Then, in section 3, we
critically assess their interpretation. Our main focus in this section is on
the meaning of information in their approach to verify that to what extent
(if any) they could coherently explain the peculiarities of quantum world.
Subsequently, in section 4, we will show that once one is going to interpret
Brukner and Zeilinger's measure of information in a real extension of the
spin measurement, it cannot quantify the amount of uncertainty in a suitable
manner.

\section{Elements of Brukner and Zeilinger's Interpretation of Quantum
Mechanics}

The first topic in Brukner and Zeilinger's interpretation is that the whole
physical description is based on \textit{propositions}. Application and
selection of propositions is not arbitrary in their attitude; rather it
depends on what we want to learn about Nature and what knowledge about
Nature we intend to discuss with others\footnote{%
This can be interpreted as an indication of \textit{intentionality} in the
transmission of information which demands a \textit{semantic} concept of
information. We will talk more about this point in next section.}.
Emphasizing the role of propositions in physical description, Brukner and
Zeilinger persist in a fundamental point which is a basic element in their
interpretation: A quantum system is a construct based on complete list of
propositions together with their truth values. A proposition, they assert
for example, could be (1) ''The velocity of the object is $v$'' or (2) ''The
position of the object is $x$''. But since in quantum mechanics it is
usually impossible to determine simultaneously the truth values of two
arbitrary propositions, if the proposition (1) is definite, then the
proposition (2) is completely indefinite and vice versa. The two
propositions, here, are mutually exclusive. Brukner and Zeilinger consider
this as a special case of quantum complementarity. In this way, a new
language is represented for the physical description of quantum formalism, a
language in which the propositions construct the \textit{essence} of quantum
systems instead of merely describing them. This is a starting point for
critical arguments about their interpretation.

The language of propositions in quantum mechanics is similar to the language
which is used for the description of classical systems, except for the
complementary situations which must also be taken into account in the
quantum description. While we can always assign definite truth values to all
propositions in (deterministic) classical theories, one cannot assign
simultaneously definite truth-values to mutually exclusive propositions in
quantum mechanics. But even in cases that the truth values of two mutually
exclusive propositions cannot be assigned simultaneously, we can always
specify the amount of information about them. So, according to Brukner and
Zeilinger, this reveals a more fundamental notion which is \textit{knowledge}
and \textit{information}: We can quantify our information about descriptive
propositions in both the classical and quantum domains. What is changed,
however, in passing from classical to quantum world is the replacement of
the definite truth values with the finite amounts of information. They
believe that this is the least expense which could be paid for the change in
the epistemological structure of classical physics. In this view,
information is regarded as \textit{the most fundamental notion} [8]:

\begin{quote}
``In contrast [with Bell's idea] it is suggested that \textit{information}
is the most basic notion of quantum mechanics, and it is \textit{information}
about possible \textit{measurement} results that is represented in the
quantum states.''\footnote{%
Emphases in all quotations throughout this text are original.}
\end{quote}

We believe, however, the expense is high, since this expense is not merely
paid for substituting the non-deterministic finite information in quantum
mechanics in place of the deterministic complete description of classical
systems. It is also paid for introducing the finiteness of information as an
indispensable ingredient of a quantum object itself. What is replaced here,
is the information content \textit{of} the systems, not merely the amount of
available information \textit{about} the systems [4]:

\begin{quote}
``\textit{The information content of a quantum system is finite.}

With this we mean that a quantum system cannot carry enough information to
provide definite answers to all questions that could be asked
experimentally. Then, by necessity the answer of the quantum system to some
questions must contain an element of randomness. This kind of randomness
must then be irreducible, that is, it cannot be reduced to `hidden'
properties of the system. Otherwise the system would carry more information
than what is available.''
\end{quote}

Subsequently, Brukner and Zeilinger conclude that irreducible randomness
results from the finiteness of information [4]:

\begin{quote}
``Thus, without any additional physical structure assumed, we let the
irreducible randomness of an individual event and complementarity [both], be
a consequence of the finiteness of information.''
\end{quote}

In response to the question that how much information a quantum system can
carry, they introduce the principle of quantization of information for an
elementary system. According to this principle, a descriptive proposition
for a complex system can be subdivided to constituent propositions until we
reach a final limit. The individual system that represents the truth value
of one single proposition only, is called elementary system. Beyond this
limit, information is irreducible. So the principle of quantization of
information states that [1]:

\begin{quote}
``\textit{An elementary system carries 1 bit of information.''}
\end{quote}

For a complex system, consisting of $N$ elementary systems this principle is
generalized to [1]:

\begin{quote}
``\textit{N elementary systems carry N bits.''}
\end{quote}

The principle of quantization of information does not make any statement
about how the $N$ bits of information are distributed over the $N$ systems.
Brukner and Zeilinger consider this principle for systems with both
independent and entangled subsystems.

For example, for a spin-$\frac{1}{2}$ system (like an electron) there are
always three complementary or mutually exclusive propositions for describing
the spin situation. These are: ''The spin along $\stackrel{\rightarrow }{%
n_{1}}$ is up (down)'', ''The spin along $\stackrel{\rightarrow }{n_{2}}$ is
up (down)'' and ''The spin along $\stackrel{\rightarrow }{n_{3}}$ is up
(down)'', where $\stackrel{\rightarrow }{n_{1}}$, $\stackrel{\rightarrow }{%
n_{2}}$ and $\stackrel{\rightarrow }{n_{3}}$ are mutually orthogonal
directions. Complete information about one of the propositions is possible
only through complete ignorance of two others. Also for any other two-valued
observable, there are three complementary propositions for describing the
state of system, even though these propositions are not related with three
space directions.

Now this question arises as to how the amount of information (or the amount
of uncertainty) could be quantified in a statistical prediction? Brukner and
Zeilinger remark that in classical measurements in which it is assumed that
the classical system has predetermined physical properties before a
measurement is performed, the Shannon information is an appropriate measure
of uncertainty which can be defined as the following relation for $n$
possible outcomes:

\begin{equation}
H(\stackrel{\rightarrow }{p})=-\sum\limits_{i=1}^{n}p_{i}\log p_{i}  \tag{1}
\end{equation}
where $\stackrel{\rightarrow }{p}$ is a probability vector in the
probability space with length $\left| \stackrel{\rightarrow }{p}\right| =%
\sqrt{\sum\limits_{i=1}^{n}p_{i}^{2}}$ in which $p_{i}$ is the probability
of occurrence of the outcome $i$ in a given measurement. For quantum
measurements, however, they introduce a new measure of information which is
defined as [2, 3]:

\begin{equation}
I(\stackrel{\rightarrow }{p})=\mathcal{N}\sum\limits_{i=1}^{n}\left( p_{i}-%
\frac{1}{n}\right) ^{2}  \tag{2}
\end{equation}
where $\mathcal{N}$ is a normalization factor. In general, for a system in
which maximally $k$ bits of information can be encoded, $n=2^{k}$ and $%
\mathcal{N}=2^{k}k\diagup \left( 2^{k}-1\right) $. For an elementary system
which can carry only one bit of information, $n=2$. Then, one can reach the
following relation:

\begin{equation}
I(p_{1},p_{2})=2(p_{1}-\frac{1}{2})^{2}+2(p_{2}-\frac{1}{2}%
)^{2}=(p_{1}-p_{2})^{2}  \tag{3}
\end{equation}
where $p_{1}$ and $p_{2}$ are the probabilities defined for a dichotomic
observable and $p_{1}+p_{2}=1$. If one of the probabilities is one, $I$
reaches its maximal value of 1 bit of information which is equivalent to the
complete certainty. If both probabilities are equal, $I$ takes its minimal
value of 0 bits of information which implies a complete uncertain (or a
complete random) situation.

We can write the measure of information (3) as $I=i^{2}$ where $%
i=p_{1}-p_{2} $. Moreover, one can generalize the relation (3), if the
vector $\stackrel{\rightarrow }{i}$ is defined in the information space as $%
\stackrel{\rightarrow }{i}=\left( i_{1},i_{2},i_{3}\right)
=(p_{x}^{+}-p_{x}^{-},p_{y}^{+}-p_{y}^{-},p_{z}^{+}-p_{z}^{-})$, where,
e.g., $p_{z}^{+}$ denotes the probability of finding the spin-up result
along the $z$-direction. Vector $\stackrel{\rightarrow }{i}$ characterizes
the information state of an elementary system which is, in Schr\"{o}dinger's
terminology, a \textit{catalog of knowledge} about a set of three mutually
complementary propositions [4]:

\begin{quote}
``It is assumed that the catalog $\stackrel{\rightarrow }{i}$ is a complete
description of the system in the sense that its knowledge is sufficient to
determine the probabilities for the outcomes of all possible future
measurements.''
\end{quote}

Then, the quantum state is also considered as the catalog of knowledge.

According to the measure of information indicated in relation (3), the total
information of an elementary system is defined as a sum of the individual
measures of information over a complete set of mutually complementary
propositions

\begin{equation}
I_{total}=I_{1}+I_{2}+I_{3}=1  \tag{4}
\end{equation}
where, $I_{1}=i_{1}^{2}$, $I_{2}=i_{2}^{2}$ and $I_{3}=i_{3}^{2}$. Since an
elementary system carries 1 bit of information, the total information is
also equal to one. So, in different experimental setups where the components
of $\stackrel{\rightarrow }{i}$ adopt different values, the total
information is always constant and equal to one for pure states.

In Brukner and Zeilinger's interpretation, an entangled state is also
important from this point of view that an entangled system has more
information than its individual constituents [1, 4, 5]. For example,
consider the following singlet state for a pair of spin-half particles which
is known as one of the four entangled Bell states:

\begin{equation}
\left| \Psi ^{-}\right\rangle =\frac{1}{\sqrt{2}}\left( \left| \widehat{n}%
,+\right\rangle _{1}\left| \widehat{n},-\right\rangle _{2}-\left| \widehat{n}%
,-\right\rangle _{1}\left| \widehat{n},+\right\rangle _{2}\right)  \tag{5}
\end{equation}
where $\left| \widehat{n},+\right\rangle _{1}$, e.g., is the spin-up state
of particle 1 along $\widehat{n}$-direction (which can be chosen arbitrary).
The total information represented by the entangled state (5) is $%
I_{Corr}^{Bell}=2$. This means that two bits of information are carried by
the whole system, corresponding to statements about the results of joint
observations, ``The two spins are the same along $x$'' and ``The two spins
are the same along $y$''. The truth values of these propositions is
false-false. Since any product state of two particles (like each single term
in relation (5)) carries only one bit of information, for any entangled
state we have $I_{Corr}^{entgl}>1$\footnote{$I_{Corr}^{entgl}$ can be
calculated for any set of spin measurements of two particles along two
arbitrary directions. For more details, see for example [7].}. This is an
important information condition which is the characteristic feature of
entangled states. Entanglement here is believed to be a secondary concept
which its advantage is only due to its information content.

In this interpretation, wave function is only a mathematical representation
which contains encoded information about a quantum system. Our knowledge
about a system gives it the possibility of being in a place at a given
moment. In the measurement process, the quantum state which represents that
knowledge changes. This change is instantaneous, but it is only a change in
our knowledge. Our information about all of the other spatial points also
changes instantaneously. However, there is nothing to be transmitted in a
physical manner: There is no faster than light signalling. In addition, the
total information content of the system remains constant [4]:

\begin{quote}
``Unlike a classical measurement, a quantum measurement thus does not just
add (if any) some knowledge, it changes our knowledge in agreement with a
fundamental finiteness of the total information content of the system.''
\end{quote}

The total information content of a system, $I_{total}$, is invariant under
the change of the representation of the catalog $\stackrel{\rightarrow }{i}$
and remains constant under any rotation in the information space. Brukner
and Zeilinger determine the rotation matrices for Euler angles, with
assumptions: (1) The invariance of the total information under rotation and
(2) the homogeneity of rotational angles (in the sense that adding a
constant value to any of three rotational directions, does not change the
physics of the problem and the location of vector $\stackrel{\rightarrow }{i}
$ in the information space remains unchanged) [4]. According to these two
assumptions, they show that for a spin-half particle, angular momentum
generates the rotation. Correspondingly, they want to extend a new
formulation of quantum mechanics which is constructed on the information
space, instead of an abstract vector space. Moreover, they show that the
following dynamical equation in quantum mechanics,

\begin{equation}
i\hbar \frac{d\widehat{\rho }(t)}{dt}=\left[ \widehat{H}(t),\widehat{\rho }%
(t)\right]  \tag{6}
\end{equation}
can be derived for an \textit{elementary system\ }from the time evolution of
catalog of knowledge $\stackrel{\rightarrow }{i}$, in the information space
[4]. In relation (6), $\widehat{\rho }(t)$ and $\widehat{H}(t)$ are the
density matrix and the Hamiltonian of system, respectively. A key assumption
here is that, for an isolated system with no information exchange with the
environment, the total information of system is conserved in time. I.e.,

\begin{equation}
I_{total}(t)=\sum\limits_{n=1}^{3}i_{n}^{2}(t)=\sum%
\limits_{n=1}^{3}i_{n}^{2}(t_{0})=I_{total}(t_{0})  \tag{7}
\end{equation}

There are still unsolved problems which are to be solved, as mentioned by
themselves too. The measure of information defined in relation (2) has not
yet been studied for continuous variables. One should also elucidate that in
the information space, e.g., how momentum can be inferred as the generator
of transformation and how the Schr\"{o}dinger equation can be reformulated
consistently. There is also no complete set of complementary observables for
infinite Hilbert spaces\footnote{%
We are grateful to anonymous referee for reminding this point.}.

To sum up, the ultimate goal in Brukner and Zeilinger approach is to show
that information is a fundamental notion which not only the inherent nature
of quantum particles and quantum processes are based on its concept, but
also a new formulation can emerge from its foundation. External reality
depends as well on our experimental answers to questions about the quantum
events [4]:

\begin{quote}
``Therefore the experience of the ultimate experimenter is a stream of
(`yes' or `no') answers to the questions posed to Nature. Any concept of an
existing reality is then a mental construction based on these answers. Of
course this does not imply that reality is no more than a pure subjective
human construct. From our observations we are able to build up objects with
a set of properties that do not change under variations of modes of
observation or description. These are `invariants' with respect to these
variations.''
\end{quote}

Note that in the above expression, what can be predicted by an observer is a 
\textit{qualitative} picture of physical properties. This point has been
indicated explicitly by Zeilinger elsewhere [8]:

\begin{quote}
``[T]he observer has a qualitative, but not a quantitative influence on
reality. She can define which quality will show up in the experiment, but
not the quantity, the exact value, the latter being completely random,
except in the rare case when the quantum system is in an eigenstate of the
observed quantity.''
\end{quote}

By this, Zeilinger means that for example when an observer wants to measure
the energy of a system, she knows that with her measurement the property of
energy will appear. She can define energy as a qualitative notion, but its
exact value is not predictable or definable.

Meanwhile, the randomness of the events arises from the fact that
information is finite [4]:

\begin{quote}
``It is beyond the scope of quantum physics to answer the question why
events happen at all (that is, why the detectors clicks at all). Yet, if
events happen, then they must happen randomly. The reason is the finiteness
of the information.''
\end{quote}

Even the Bohr complementarity is regarded as a consequence of the finiteness
of information [8]:

\begin{quote}
``Complementarity then simply is a consequence of the fact that the total
information which is represented by a quantum system is finite.''
\end{quote}

\section{Comments On The Information Concept}

In reformulating the quantum dynamical equation (6), two basic assumptions
were considered: 1) The total information is a conserved quantity and 2) the
components of the information vector generates SU(2). Second assumption says
that the time evolution of the information vector is unitary. Since in this
interpretation, we are presented with the variation of the information
vector together with the invariance of the total information in the \textit{%
measurement} process too, this important question appears as to why it is
not defined any time evolution for the information vector during the
measurement. Note that this question comes within the scope of the
interpretation itself and does not refer to subjects like the objectivity of
the quantum system. So, it seems that there are two kinds of information
debated here: Information before measurement and information after
measurement. The time evolution of information before measurement is unitary
and reversible with time, but information during measurement evolves
irreversibly: We can always reproduce the initial information from the
information given at a later time (and vice versa) except for the case that
a measurement is performed in the intervening time. Why should the
information vector evolve in different ways? What is the concept of
instantaneous change in information that occurs during the measurement? Does
information really have a dual nature? How does the finiteness of
information explain this duality? If information is a fundamental concept,
in what manner can one realize the distinction in the basis?

On the other hand, many people think that quantum mechanics must agree with
classical mechanics in a suitable limit. So, if quantum dynamics could be
constructed on an underlying definition of information, it would be expected
that in such an approach, the new measure of information could become
consistent with a classical description in an appropriate limit too.
Accordingly, what is the relation between the classical and the quantum
worlds as far as the notion of information is concerned? And considering
Brukner and Zeilinger's new measure of information, how can one explain the
apparent consistency of the two worlds in Nature?

In this regard, Timpson distinguishes two points of view in Brukner and
Zeilinger interpretation: an instrumentalism versus a phenomenalism [12,
13]. In this interpretation, a quantum state merely describes the
probabilities of the possible measurement results, a way for representing
the outcomes of all possible future observations [1]. This characterizes
their instrumentalist point of view. On the other hand one can discern a
form of phenomenalism according to which physical objects are taken to be
not an actual things, but to be some constructs relating observations. So,
as Timpson states [12],

\begin{quote}
``[A] system represents a quantity of information about measurement results
because a physical system literally \textit{is} nothing more than an
agglomeration of actual and possible sense impressions arising from
observations.''
\end{quote}

As a consequence of the combination of these two attitudes, Timpson argues
that the principle of quantization of information cannot function as a
foundational principle for quantum mechanics, nor does it have the
explanatory power that Brukner and Zeilinger suppose [12, 13].

It is, however, important to remember that all features of this
interpretation cannot be categorized as instrumentalist or phenomenalist
perspectives. As an instance, one can mention the very meaning of
information here which is introduced too fundamentally to be understood as a
form of phenomenalism or instrumentalism. Nonetheless, one sees no clear
definition of information and its conception in this interpretation. (A
detailed discussion of this subject is given in Appendix.)

The principle of finiteness of information is also a matter of controversy.
Indeed, how can one verify that a quantum system must essentially have a
finite information content? Many people may still be interested to know why
we should have such a constraint about the possible measurement results. As
many people, for example, are still curious to know why the thermodynamic
systems inherently tend to go towards most probable states according to the
Second Law of Thermodynamics. Since, one may some day discover a
thermodynamic force which will explain the underlying reason of this
tendency.

Yet, there are some people who believe that an \textit{unlimited} amount of
information can be coded in a given quantum state. This can precisely be
illustrated by Wiesner's quantum multiplexing argument which states that one
can always code two distinct one-bit messages into a spin-$\frac{1}{2}$
system [17]. However, the spin state is not an observable and there is no 
\textit{accessible} information larger than a single bit. No observer can
read both bits, but it is only possible to read one bit of information in
any proper choice of measurement. The finiteness of gaining knowledge about
the properties of a system is not a specific feature of quantum mechanics
only. Information constraints also exist for classical systems [18]. On the
other hand, any argument about the finiteness of information is contingent
upon the fact that how one interprets the probability, not upon the fact
that how one determines the quantity of information.

Brukner and Zeilinger consider the randomness as an intrinsic property for
quantum systems. But, they conclude this not as a possible interpretation of
probability coming from outside the theory. Instead, they believe that this
is a natural consequence of the principle of finiteness of information. But,
as Markus Arndt has recently noted [19]:

\begin{quote}
``Objective (absolute) randomness is hardly fully verifiable. One may
exclude certain classes of causes / reasons for a quantum choice .........
But the a priori exclusion of any reason whatsoever cannot be falsified /
verified.''
\end{quote}

Brukner and Zeilinger try to show that everything, even an objective event
which happens in an experiment results from the finiteness of information.
This point is expressed explicitly in the abstract of one of their papers in
which they discuss conceptually the double-slit experiment for $C_{60}$
molecules [6]:

\begin{quote}
``It is argued here that quantum interference is a consequence of the
finiteness of information.''
\end{quote}

This attitude originates again from the fact that information plays a
central role in quantum world. So, they attempt to demonstrate that
objective properties of quantum systems which appear in measurements are
merely informational effects. Subsequently, they conclude that the
interference pattern of quantum particles is an objective consequence of the
finiteness of information. In other respects, information loses its
fundamental role and the question that how the interference pattern is
formed (or any other objective property that appears in a measurement)
remains without answer.

Nevertheless, what really they show is that:

\begin{quotation}
$1$) there exists a limitation in observing both the path and the
interference pattern of a quantum particle simultaneously and,

$2$) this limitation is a consequence of the finiteness of information
content of a quantum system.
\end{quotation}

In other words, what they show is that we should not expect we could obtain
a desirable classical knowledge about a quantum object. Logically, however,
this does not mean that:

\begin{quotation}
$2^{\prime }$) the interference too, is a consequence of our constraints in
acquiring the information.
\end{quotation}

There is a clear distinction in the meaning of statements $2$ and $2^{\prime
}$. The statement $2^{\prime }$ cannot be deduced naturally from their
discussion, but (it seems) it relies primarily on their phenomenalistic
position about the microobjects. This point of view is not of course
compelling. But, even if we adopt this view, we cannot still conclude that
the interference pattern is a result of the finiteness of information too.
The finiteness of information does not explain \textit{how} these objective
events appear, even though we put it into the ontology of quantum objects.

At the end, it must be pointed out that some kind of incongruity is realized
in their interpretation. In discussing the objective reality and its
relation with information, at first sight, Zeilinger takes these notions on
the same footing, so that neither one is sufficient for understanding the
quantum world [1]:

\begin{quote}
``Therefore, while in a classical worldview, reality is a primary concept
prior to and independent of observation with all its properties, in the
emerging view of quantum mechanics the notions of reality and of information
are on an equal footing. One implies the other and neither one is sufficient
to obtain a complete understanding of the world.''
\end{quote}

But after that, Brukner and Zeilinger take stronger position and consider
the physical properties of objects as secondary and information as primary
notions. In other words, a real property is viewed as a \textit{%
representation} of information which is \textit{created} in measurement [4]:

\begin{quote}
``In classical physics a property of a system is a primary concept prior to
and independent of observation and information is a secondary concept which
measures our ignorance about properties of the system. In contrast in
quantum physics the notion of the total information of the system emerges as
a primary concept, independent of the particular complete set of
complementary experimental procedures the observer might choose, and a
property becomes a secondary concept, a specific representation of the
information of the system that is created spontaneously in the measurement
itself.''
\end{quote}

There is no further explanation that \textit{why} the reality of physical
properties becomes suddenly a secondary concept. Also, it is not obvious 
\textit{how} reality is created spontaneously in the measurement.

\section{Critical Assessment of Brukner and Zeilinger's measure of
information}

Brukner and Zeilinger's definition of a measure of information is derived
initially from an \textit{uncertainty} expression for a specific outcome in $%
N$ trials of an experiment. This can be defined as $\frac{\sigma ^{2}}{N}%
=p(1-p)$ where $p$ is the probability of the occurrence of a dichotomic
result and $\sigma ^{2}$ is variance for binomial distribution [2, 4]. Then,
for $n$ outcomes with the probabilities $\stackrel{\rightarrow }{p}\equiv
(p_{1},p_{2},....,p_{n})$ of the individual occurrences, they deduce the
relation (2) as an appropriate measure of information.

A crucial point here is that their measure of information requires not only
that probabilities are known, but also that the number of possible outcomes $%
n$ is known (see relation (2))\footnote{%
We are grateful to anonymous referee for emphasizing this point.}. In turn,
the value of $n$ is determined by $k$ which is defined as the maximum bits
of information that can be encoded in a quantum system [2, 4]. As a matter
of fact, this means that two different notions of total information are to
be taken into account in Brukner and Zeilinger's interpretation:

1) A total information $k$ defined as an input of the relation (2) which is
determined by knowing the number of possible outcomes $n$,

2) A total information $I_{tot}$ which is an output of relation (2) as the
sum of the individual measures of information.

Considering those cases where $k$ bits of information can be maximally
encoded, Brukner and Zeilinger define $n=2^{k}$, or $k=\log _{2}n$ [2]. But,
where is basically $k$ coming from? The answer is that if one supposes $n$
equiprobable events which each event could occur with the same probability $%
\dfrac{1}{n}$, $k$ would be precisely the Shannon information $H(\stackrel{%
\rightarrow }{p})$ for $p_{i}=\dfrac{1}{n}$. Here, $k$ is essentially
endowed with a counterfactual meaning of information in the Shannon sense of
definition. One may consequently interpret that Brukner and Zeilinger's
measure of information in (2) does need a form of the so-called
microcanonical Shannon entropy as an input. So, $k$ and $I_{tot}$ are not
principally the same and it is expected that they should not be the same
numerically in some circumstances too.

To elaborate this point more palpably, let us consider that in a \textit{real%
} experiment, there will be a non zero probability of non-detection for
measuring each spin component of a spin half particle. Suppose also that, 
\textit{ideally}, the spin of particle would be expected to be up along $x$.
Then, if the overall efficiencies of the measuring apparatuses are assumed
to be the same along different directions, one can describe the statistics
of the three possible events of spin up detection ($+$), spin down detection
($-$) and no detection ($0$) along $x$, $y$ and $z$ with the following
probabilities, respectively:

\begin{eqnarray}
p_{x}^{+} &=&\eta ,\text{ }p_{x}^{-}=0,\text{ }p_{x}^{0}=1-\eta ;  \nonumber
\\
\text{ }p_{y}^{+} &=&p_{z}^{+}=\frac{1}{2}\eta ,\text{ }p_{y}^{-}=p_{z}^{-}=%
\frac{1}{2}\eta ,\text{ }p_{y}^{0}=p_{z}^{0}=1-\eta  \tag{8}
\end{eqnarray}
where $\eta $ denotes the overall efficiency; $p_{x}^{0}$, e.g., is the
probability of non-detection along $x$ and the detection probabilities are
defined as before. Using the relation (2), one can obtain the individual
information functions $I_{1}$, $I_{2}$ and $I_{3}$ along $x$, $y$ and $z$,
respectively, for $n=3$ and $k=\log _{2}3$:

\begin{eqnarray}
I_{1} &=&\dfrac{3\log _{2}3}{2}\left[ \left( \eta -\frac{1}{3}\right)
^{2}+\left( \eta -\frac{2}{3}\right) ^{2}+\frac{1}{9}\right] ;  \nonumber \\
I_{2} &=&I_{3}=\dfrac{3\log _{2}3}{2}\left[ \frac{1}{2}\left( \eta -\frac{2}{%
3}\right) ^{2}+\left( \eta -\frac{2}{3}\right) ^{2}\right]  \tag{9}
\end{eqnarray}
After some simple algebra, one can get

\begin{equation}
I_{total}=I_{1}+I_{2}+I_{3}=\dfrac{3\log _{2}3}{2}\left( 5\eta ^{2}-6\eta
+2\right)  \tag{10}
\end{equation}

The ratio of $\dfrac{I_{total}}{k}$ plotted as a function of $\eta $ is
shown in figure1. For $\eta \rightarrow 1$ (perfect efficiency), the number
of possible outcomes $n$ reduces to $2$ (i.e., there is no non-detection
event) and the relation (3) will be the appropriate expression for each
individual measure of information. Then, $I_{total}=k=1$ describes the total
information content of an elementary system (in our example, a spin half
particle), according to Brukner and Zeilinger's interpretation. However, for 
$0<\eta <1$, there are three possible outcomes and while in such an
experiment $k$ denotes that a value of $\log _{2}3\approx 1.585$ bits of
information can be maximally encoded (according to their definition), $%
I_{total}$ is given by (10) which depends on the efficiency factors of
measuring devices. Since, in different experiments $\eta $ may not be the
same, $I_{total}$ cannot be considered as a \textit{conserved} value in
principle. Specially, for $0<\eta <0.29$ and $0.91<\eta <1$, $I_{total}$
exceeds the maximal possible amount of information, i.e., $1.585$ bits that
can be encoded in an individual spin half particle. This is in contrast to
what is expected for $I_{total}$ in the Brukner and Zeilinger's sense of
definition [3]:

\begin{quote}
``Independent of the various possibilities to encode information, the total
information content of the system cannot fundamentally exceed the maximal
possible amount of information that can be encoded in an individual
observable.''
\end{quote}

In contrast, Shannon information calls only for a knowledge of probabilities
and its change is isomorphic for different $n$. Its description is also
clear for the above example. Let us define $H_{x}$, $H_{y}$ and $H_{z}$ as
the corresponding Shannon uncertainties along the directions $x$, $y$ and $z$%
, respectively. Then, one can obtain:

\begin{eqnarray}
H_{x} &=&-\eta \log _{2}\eta -(1-\eta )\log _{2}(1-\eta );  \nonumber \\
H_{y} &=&H_{z}=-\eta \log _{2}\dfrac{\eta }{2}-(1-\eta )\log _{2}(1-\eta ) 
\nonumber \\
&=&H_{x}+\eta  \tag{11}
\end{eqnarray}
While for $\eta \rightarrow 1$, $H_{x}$ indicates that there is no
uncertainty about the value of the spin component of particle along $x$, it
increases as $\eta $ is being reduced to $\dfrac{1}{2}$. This is the point
where we have the maximum uncertainty of one bit for the spin component
along $x$. Moreover, $H_{y}$($=H_{z}$) reaches its maximum value of $\log
_{2}3\approx 1.585$ bits (which is the same value of $k$) for $\eta =\dfrac{2%
}{3}$. Nevertheless, as $\eta $ approaches to zero, all the uncertainties $%
H_{x}$, $H_{y}$ and $H_{z}$ decrease, because the probabilities of detection
along different directions approaches to zero. The uncertainties $H_{x}$ and 
$H_{y}$($=H_{z}$) are plotted against $\eta $ in figure 2.

The above example shows that the Shannon and the Brukner-Zeilinger measures
of information may have very different descriptions even for simple cases,
when merely the primary definitions are extended to more concrete
circumstances. It casts doubt upon the fact that whether the latter measure
of information can retain its significance for more realizable physical
examples.

\section{Conclusions}

Brukner and Zeilinger's interpretation has brought up some provoking topics.
Statements about the essential nature of information, the finiteness of the
information content of a quantum object, the possibility of deriving the
quantum dynamics from an information basis and the definition of a new
measure of information, all are interesting subjects which call for careful
analysis. Considering information as a fundamental notion, however, this
interpretation encounters some difficulties and leaves us alone with not
answered basic questions. It is not clear why a quantum object should have a
finite information content in its essence, how the information vector evolve
when one measures a physical quantity, and what really means as an
instantaneous change in information during the measurement. Here, we are
confronted with a complex, multidimensional notion of information which its
foundation is not clear.

On the other hand, Brukner and Zeilinger's definition of a new measure of
information may lose its significance, when an elementary system is treated
realistically in the spin measurements. Here, the possible number of
outcomes $n$ may discriminate the value of the total information content in
ideal and non-ideal measurements. So, one is encountered this basic question
that what does an elementary system actually mean, when its information
content cannot be conserved in real practice?

\textbf{Acknowledgment. }A. Shafiee acknowledges Atomic Energy Organization
of Iran (AEOI) for financial support. The authors also thank anonymous
referee of this journal for many helpful comments.\bigskip

\-\textbf{Appendix\medskip }\newline
Information is a notion with different concepts, and each concept has its
own application(s) [20, 21, 22]. It is hardly possible that a single concept
of information can account for the various applications of this extensive
field. However, it is worth looking into a special notion of information
here: Information with \textit{semantic} concept is something capable of
providing knowledge [20]. In Routledge Encyclopedia of Philosophy, one reads
[23] ``The term information has various senses in ordinary discourse,
including knowledge ..... and propositional content ..... Information in
these senses has semantic features, such as reference and truth-value.''

The main aspect of semantic information is about what a message or
proposition means. In this sense, knowledge and semantic information are
relative notions. To have knowledge about a system, we should describe the
information content of that system by using the descriptive propositions.
Once information is available, knowledge can be built in terms of justified
information.

As stated by Dretske in his semantic theory of information (see [24], p.
45): ``A state of affairs contains information about X to just that extent
to which a suitable placed observer could learn something about X by
consulting it'' (Quoted also in [20]). Dretske also qualifies information as
a semantic concept in virtue of the intentionality inherent in its
transmission. Information receives its intentional character from the lawful
regularities on which it depends (for example, the laws of physics in the
domain of physics). Yet, it is not relevant in this view to ask where the
intentional character of laws comes from (see [24], p. 77 and also [20]).
This makes some conceptual difficulties regarding a well-defined relation
between information and the physics from which information stems. Floridi
also analyzes the meaning of semantic information in terms of well-formed,
meaningful and truthful data with a propositional orientation [22]. Here,
``well-formed'' means that data have been put together correctly according
to some given rules and ``meaningful'' means that the data must also comply
with the meanings of propositions. So, the semantic content in his view is
formed on the basis of the well-formed and meaningful data represented by
propositions with definite truth values which in turn are used to talk about
the world and describe it.

On the other hand, as Shannon himself remarks [9]:

\begin{quote}
``[S]emantic aspects of communication are irrelevant to the engineering
problem. The significant aspect is that the actual message is one selected
from a set of possible messages.''
\end{quote}

The Shannon measure of information is purely quantitative. It only deals
with amounts of information and disregards questions related to information
content or the meaning of information. Shannon information is commonly
described as a study of information at the \textit{syntactic} level, which
can be studied by the Mathematical Theory of Communication, also known as
Communication Theory or Information Theory. According to syntactic concept,
information theory is introduced with the application of random variables
and probability distributions, so it is a mathematical theory. It has
applicability to a variety of fields like communication theory,
thermodynamics, computer science and algorithmic complexity [20].

The concept of information can also be treated physically. By the term 
\textit{physical}, we mean that information either can be transmitted via a
physical entity (i.e., a signal), or is an objective property of the system
which can be described by the physical laws, or both. According to this
view, information is transmitted from one point to another point via a
physical entity, i.e., an information-bearing signal. This is in opposition
to a semantic view where states that to establish an informational link
between two distant locations, it is only necessary that one can have a
knowledge at one location by looking at the other one. This point is similar
to what is declared by Brukner and Zeilinger in one of their articles [4]:

\begin{quote}
``When the state of a quantum system has a non-zero value at some position
in space at some particular time, it does not mean that the system is
physically present at that point, but only that our knowledge (or lack of
knowledge) of the system allows the particle the \textit{possibility} of
being present at that point at that instant.''
\end{quote}

There is no direct reference to a specific notion of information in Brukner
and Zeilinger's interpretation. Nonetheless, it seems that the meaning of
information has a complex character in their approach. On one side, it
appears that a semantic character for information is more justified here.
For a given system, measure of information implies what can be said
quantitatively about possible measurement results by assigning truth values
to corresponding descriptive propositions. The quantum state also represents
the possible ways one can assign truth values to these propositions.

On the other side, information is treated as \textit{the} base for quantum
physics in this interpretation. So, how can one reformulate the quantum
dynamics with a notion of information which seems to be not physical in its
essence? Hence, the question appears as to whether quantum physics is
reducible to an information modeling. Could the universe primarily be\textit{%
\ made} of information? And if so, how does matter emerge from information?

\end{document}